\documentstyle[epsfig]{article}

\def\abstract#1{\vskip 7mm
        \begin{center}{\large Abstract}\par \smallskip
                \begin{minipage}[c]{12cm}
                        \small #1
                \end{minipage}
        \end{center}
}
\def\title#1{\begin{center}{\Large\bf #1}\end{center}}
\def\author#1{\vskip 5mm \begin{center}{#1}\end{center}}
\def\address#1{\begin{center}{\it #1}\end{center}}
\makeatletter
\@ifundefined{lesssim}{}{}
\@ifundefined{gtrsim}{}{}
\def\vereq#1#2{\lower3pt\vbox{\baselineskip1.5pt \lineskip1.5pt
\ialign{$\m@th#1\hfill##\hfil$\crcr#2\crcr\sim\crcr}}}
\makeatother

\begin{document}

\title{%
 Deformed phase space and canonical quantum cosmology
\smallskip \\
{\large}}
\author{%
Babak Vakili\footnote{E-mail: bvakili45@gmail.com, b-vakili@sbu.ac.ir}}
\address{%
Department of Physics, Azad University of Chalous, P. O. Box 46615-397, Chalous, Iran}
\abstract{The effects of noncommutativity and deformed Heisenberg algebra on the evolution
of a two dimensional minisuperspace quantum cosmological model are investigated.}

\section{Introduction}
We study the effects of noncommutativity and deformed Heisenberg
algebra on the evolution of a two dimensional minisuperspace
quantum cosmological model. The phase space
variables turn out to correspond to the scale factor and the anisotropic parameter of a magnetized Bianchi type I
model with a cosmological constant. The exact quantum solutions in commutative and noncommutative cases are
presented. We also obtain some approximate analytical solutions for
the corresponding quantum cosmology in the presence of
the deformed Heisenberg relations between the phase space variables,
in the limit where the minisuperspace variables are small. These
results are compared with the standard commutative and
noncommutative cases and similarities and differences of these
solutions are discussed.
\section{The model}
Let us consider a cosmological model in which the spacetime
is assumed to be of Bianchi type I whose metric can be
written, working in units where $c =\hbar = 16\pi G = 1$, as
\begin{equation}\label{A}
ds^2=-N^2(t)dt^2+e^{2u(t)}e^{2\beta_{ij}(t)}dx^idx^j,\end{equation}
where $N(t)$ is the lapse function, $e^u(t)$ is the scale factor of the
universe and $\beta_{ij}(t)$ determine the anisotropic parameters $v(t)$
and $w(t)$ as follows
\begin{equation}\label{B}
\beta_{ij}=\mbox{diag}\left(v+\sqrt{3}w,v-\sqrt{3}w,-2v\right).\end{equation}To simplify the model we take $w = 0$. The Lagrangian in the minisuperspace
$\{u,v\}$ can be written as \cite{ref1}
\begin{equation}\label{D}
{\cal
L}=\frac{6e^{3u}}{N}\left(-\dot{u}^2+\dot{v}^2\right)-\Lambda N
e^{3u}.\end{equation}Thus, with the choice of the harmonic time
gauge $N=e^{3u}$, the Hamiltonian takes the form
\begin{equation}\label{E}
{\cal H}=\frac{1}{24}\left(-p_u^2+p_v^2\right)+\Lambda
e^{6u},\end{equation}where $\Lambda$ is a cosmological constant and $p_u$ and $p_v$ denote the momenta conjugate to $u$ and $v$ respectively.
\section{Quantization of the model}
Let us now proceed to quantize the model. In ordinary
canonical quantum cosmology, use of the usual commutation
relations $[x_i,p_j]=i\delta_{ij}$, , results in the well-known representation
$p_i=-i\partial/\partial x_i$, from which the WD equation can be constructed as
\begin{equation}\label{F}
\left[\frac{\partial^2}{\partial u^2}-\frac{\partial^2}{\partial
v^2}+24\Lambda
e^{6u}\right]\Psi(u,v)=0.
\end{equation}It is easy to obtain the eigenfunctions of the above equation in terms of Bessel functions as
\begin{equation}\label{G}
\Psi_{\nu}(u,v)=e^{-3\nu v}J_{\nu}\left(2\sqrt{\frac{2\Lambda}{3}}e^{3u}\right),\hspace{.5cm}\Lambda>0,\hspace{0.5cm}\Psi_{\nu}(u,v)=e^{3i\nu v}K_{i\nu}\left(2\sqrt{\frac{2|\Lambda|}{3}}e^{3u}\right),\hspace{.5cm}\Lambda<0,
\end{equation}
where $\nu$ is a separation constant. We may now write the general
solutions to the WD equations as a superposition of the
eigenfunctions
\begin{equation}\label{I}
\Psi(u,v)=\int_{-\infty}^{+\infty}C(\nu)\Psi_{\nu}(u,v)d\nu,
\end{equation}
where $C(\nu)$ can be chosen as a shifted Gaussian weight function
$e^{-a(\nu-b)^2}$.

Let us now concentrate on the noncommutativity concepts in this cosmological model.
Noncommutativity in physics is described by a deformed product, also known
as the Moyal product law between two arbitrary functions of position and momenta as \cite{ref2}
\begin{equation}\label{J}
(f*_{\alpha}g)(x)=\exp\left[\frac{i}{2}\alpha^{ab}\partial^{(1)}_a
\partial^{(2)}_b\right]f(x_1)g(x_2)|_{x_1=x_2=x},\end{equation}such
that
\begin{equation}\label{K}
\alpha_{ab}=\left(%
\begin{array}{cc}
\theta_{ij} & \delta_{ij}+\sigma_{ij} \\
-\delta_{ij}-\sigma_{ij} & \beta_{ij} \\
\end{array}%
\right),\end{equation}where the $N \times N$ matrices $\theta$ and $\beta$ are assumed to be antisymmetric with $2N$ being the
dimension of the phase space and represent the noncommutativity in coordinates and
momenta, respectively. With this product law, the deformed commutators can be written
\begin{equation}\label{L}
\left[f,g\right]_\alpha=f*_\alpha g-g*_\alpha f.\end{equation} A simple
calculation shows that
\begin{equation}\label{M}
\left[x_i,x_j\right]_\alpha=i\theta_{ij},\hspace{.5cm}\left[x_i,p_j\right]_\alpha=i(\delta_{ij}+\sigma_{ij}),\hspace{.5cm}
\left[p_i,p_j\right]_\alpha=i\beta_{ij}.
\end{equation}
Here we consider a noncommutative phase space in which $\beta_{ij} = 0$ and so $\sigma_{ij} = 0$,
i.e. the commutators of the phase-space variables are as follows
\begin{eqnarray}\label{N}
\left[u_{nc},v_{nc}\right]=i\theta,\hspace{.5cm}
\left[x_{inc},p_{jnc}\right]=i\delta_{ij},\hspace{.5cm}\left[p_{inc},p_{jnc}\right]=0.\end{eqnarray}With the noncommutative phase space defined above, we consider the Hamiltonian of the
noncommutative model as having the same functional form as equation (\ref{E}), but in which the
dynamical variables satisfy the above-deformed commutation relations, that is
\begin{equation}\label{O}
{\cal
H}_{nc}=\frac{1}{24}\left(-p_{u_{nc}}^2+p_{v_{nc}}^2\right)+\Lambda
e^{6u_{nc}}.
\end{equation}The corresponding WD equation can be obtained by the modification of the operator product
in (\ref{F}) with the Moyal deformed product
\begin{equation}\label{P}
\left[-p_u^2+p_v^2+24\Lambda e^{6u}\right]*\Psi(u,v)=0.
\end{equation}Using the definition of Moyal product (\ref{J}), it is easy to
show that
\begin{equation}\label{Q}
f(u,v)*\Psi(u,v)=f(u_{nc},v_{nc})\Psi(u,v),
\end{equation}where the relations between the noncommutative variables
$u_{nc},v_{nc}$ and commutative variables $u,v$ are given
by
\begin{eqnarray}\label{R}
p_{u_{nc}}=p_u,\hspace{.5cm}p_{v_{nc}}=p_v,\hspace{.5cm}u_{nc}=u-\frac{1}{2}\theta p_v,\hspace{.5cm}v_{nc}=v+\frac{1}{2}\theta p_u,
\end{eqnarray}Therefore, the noncommutative version of the WD
equation can be written as
\begin{equation}\label{S}
\left[\frac{\partial ^2}{\partial u^2}-\frac{\partial^2}{\partial
v^2}+24\Lambda
e^{6(u-\frac{1}{2}\theta p_v)}\right]\Psi(u,v)=0.
\end{equation}We separate the solutions into the form
$\Psi(u,v)=e^{3i\nu v}U(u)$. Noting that
\begin{eqnarray}\label{T}
 e^{6(u-\frac{1}{2}\theta p_v)}\Psi(u,v)&=&e^{6u}\Psi(u,v+3i\theta)\nonumber\\
 &=&e^{6u}U(u)e^{3i\nu(v+3i\theta)}\nonumber \\
 &=&e^{6u}e^{-9\nu \theta}\Psi(u,v),
\end{eqnarray}equation (\ref{S}) takes the form
\begin{equation}\label{U}
\left[\frac{\partial ^2}{\partial u^2}-\frac{\partial^2}{\partial
v^2}+24\Lambda e^{6u}e^{-9\nu
\theta}\right]\Psi(u,v)=0.
\end{equation}The eigenfunctions of the above equation can be written as
\begin{equation}\label{V}
\Psi_{\nu}(u,v)=e^{-3\nu v}J_{\nu}\left(2\sqrt{\frac{2\Lambda}{3}}e^{3(u-\frac{3}{2}\nu
\theta)}\right),\hspace{.5cm}\Lambda>0,\hspace{0.5cm}\Psi_{\nu}(u,v)=e^{3i\nu v}K_{i\nu}\left(2\sqrt{\frac{2|\Lambda|}{3}}e^{3(u-\frac{3}{2}\nu
\theta)}\right),\hspace{.5cm}\Lambda<0.\end{equation}
The general solutions of equation (\ref{U}) may be find as a superposition of these eigenfunctions like expression (\ref{I}).

A general predictions of any quantum theory of gravity is
that there exists a minimal length below which no other length
can be observed. An important feature of the existence of a minimal length is
the modification of the standard Heisenberg commutation relation
in the usual quantum mechanics \cite{ref3}. Such relations are
known as the Generalized Uncertainty Principle (GUP). In one
dimension, the simplest form of such relations can be written as
\begin{equation}\label{X} \left[x,p\right]=i\hbar \left(1+\beta
p^2\right),\end{equation}where $\beta$ is
positive and independent of $\bigtriangleup x$ and $\bigtriangleup
p$, but may in general depend on the expectation values $<x>$ and
$<p>$. In more than one dimensions a natural
generalization of (\ref{X}) is defined by the following
commutation relations \cite{ref3}
\begin{equation}\label{Y}
\left[x_i,p_j\right]=i\left(\delta_{ij}+\beta
\delta_{ij}p^2+2\beta p_ip_j\right).\end{equation} Now, it is easy
to show that the following representation of momentum operator in
position space fulfills the relations (\ref{Y})
in the first order in $\beta$
\begin{equation}\label{Z}
p_i=-i\left(1-\frac{\beta}{3}\frac{\partial^2}{\partial
x_i^2}\right)\frac{\partial}{\partial x_i}.\end{equation}Let us focus attention on the study of the quantum cosmology
of the model described by the Hamiltonian (\ref{E}) in GUP framework. With representation (\ref{Z}) the corresponding WD equation reads
\begin{equation}\label{AB}
\left\{\frac{\partial^2}{\partial u^2}-\frac{2}{3}\beta
\frac{\partial^4}{\partial u^4}-\frac{\partial^2}{\partial
v^2}+\frac{2}{3}\beta \frac{\partial^4}{\partial v^4}+24\Lambda
e^{6u}\right\}\Psi(u,v)=0.\end{equation} Taking $\beta=0$ in this equation
yields the ordinary WD equation where its solutions are given in equation (\ref{G}). In the case
when $\beta\neq 0$, we cannot solve equation (\ref{AB}) exactly,
but we can provide an approximate method which in its validity
domain we need to solve a second order differential equation. To
this end, note that the effects of $\beta$ are important in Planck
scales, {\it i. e.} in cosmology language in the very early
universe, that is, when the scale factor is small,
$e^{3u}\rightarrow 0$. Thus if we use the solutions (\ref{G}) in
the $\beta$- term of (\ref{AB}), we may obtain some approximate
analytical solutions in the region $e^{3u}\rightarrow 0$. The results up to first order in $\beta$ are
\begin{equation}\label{AD}
\Psi_{\nu}(u,v)=e^{-3\nu (1+3\beta \nu^2)v}J_{\nu (1+3\beta
\nu^2)}\left(2\sqrt{\frac{2\Lambda}{3}}e^{3u}\right),\hspace{.5cm}\Lambda
>0,\end{equation}
\begin{equation}\label{AE}
\Psi_{\nu}(u,v)=e^{3i\nu (1-3\beta \nu^2)v}K_{i\nu (1-3\beta
\nu^2)}\left(2\sqrt{\frac{2|\Lambda|}{3}}e^{3u}\right),\hspace{.5cm}\Lambda
<0.\end{equation}Again, the general solutions can be written as a superposition of these eigenfunctions.
\section{Summary and comparison of the results}
\begin{figure}
\begin{tabular}{cccc} \epsfig{figure=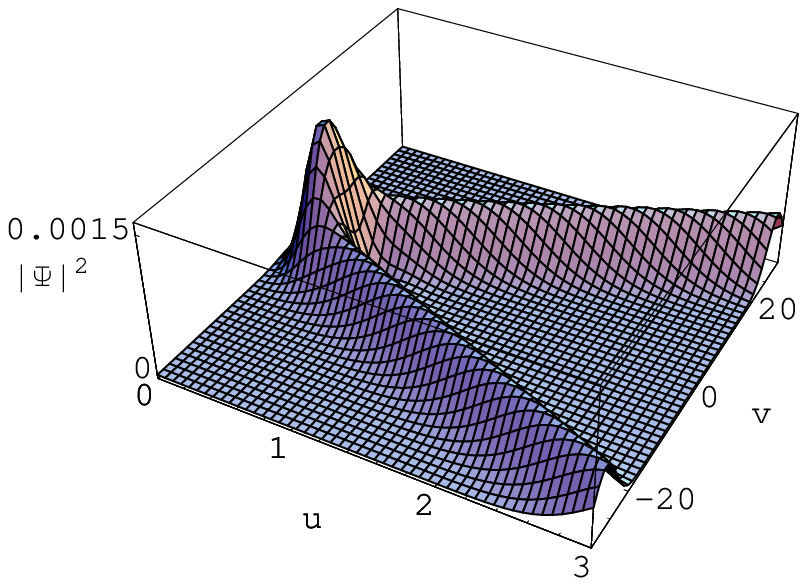,width=5cm}
\hspace{1cm} \epsfig{figure=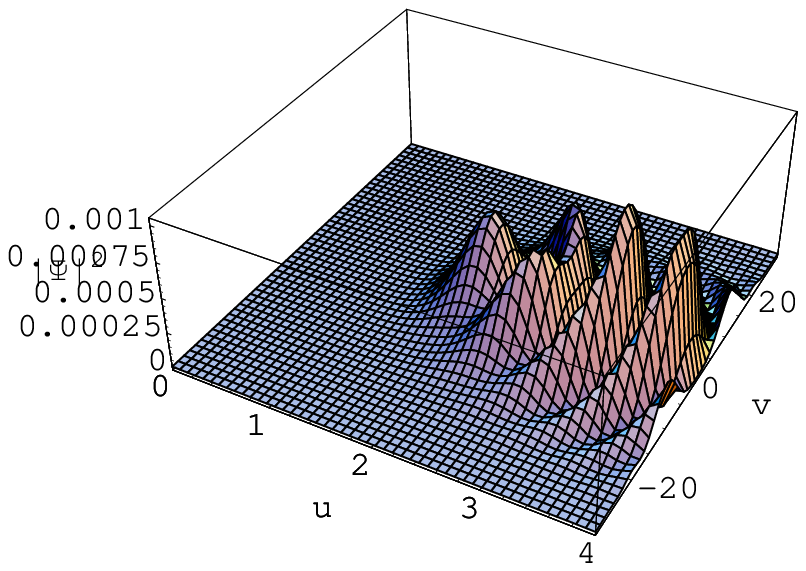,width=5cm}\hspace{1cm}\epsfig{figure=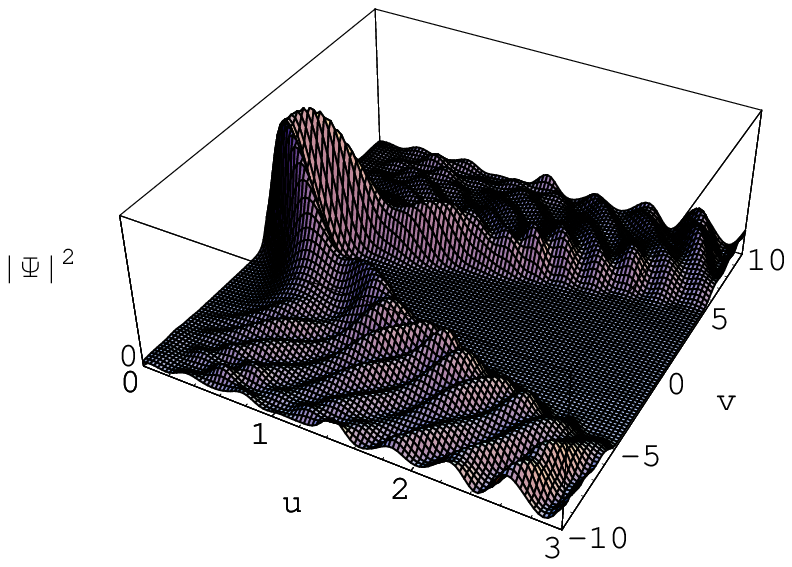,width=5cm}
\end{tabular}
\caption{\footnotesize From left to right, the figures show the square of
the commutative, noncommutative and GUP wave function. The figures are
plotted in the case of a negative cosmological constant.}
\label{fig1}
\end{figure}
In general, one of the most important features in quantum cosmology
is the recovery of classical cosmology from the corresponding
quantum model, or in other words, how can the WD wavefunctions
predict a classical universe. In figure \ref{fig1} we have plotted the square of the wave functions obtained in the previous section. As we can see from this figure, in the commutative case the peaks follow a
path which can be interpreted as the classical trajectories. The crests are symmetrically distributed around $v=0$
which may correspond to the different classical paths. Thus, it is seen
that there is an almost good correlations between the quantum
patterns and classical trajectories in the $u-v$ plane. In this case we have only one possible universe
around a nonzero value of $u$ and $v = 0$, which means that the
universe in this case approaches a flat FRW one. On the other
hand, we see that noncommutativity causes a shift in the minimum
value of $u$ corresponding to the spatial volume. The emergence
of new peaks in the noncommutative wave packet may
be interpreted as a representation of different quantum states
that may communicate with each other through tunneling. This
means that there are different possible universes (states) from
which our present universe could have evolved and tunneled,
from one state to another. Therefore, the noncommutative wavefunction predicts the emergence of the universe from a state corresponding to one of the peaks. We see that the correlation with classical trajectories is missed, i.e. the noncommutativity
implies that the universe escapes the classical trajectories and approaches a stationary state. Finally, in the GUP case as is clear from the figure the wave function has
a single peak. Although there are some small peaks in this figure, as $u$ and $v$ grow, their amplitude are
suppressed. Compare to the commutative
wave function, here we have no wave packet with peaks following the classical trajectories.
We see that instead of a series of peaks in the ordinary WD approach, we have only a single
dominant peak. This means that, similar to the noncommutative case and within the context of the
GUP framework, the wave function also shows a stationary behavior. One may then conclude that
from the point of view adopted here, noncommutativity and GUP may have close relations with each
other. However, there is an important difference, namely, that the noncommutative
wave function not only peaks around $v = 0$, but
appear symmetrically around a nonzero value of $v$, which is the
characteristic of an anisotropic universe. On the other hand, the
GUP wave function as is seen in the figure, has many peaks
around the value $v = 0$ and therefore from the point of predicting
an isotropic universe the GUP wave packet behaves like the
ordinary commutative case.

\end{document}